\documentclass{aa}
\usepackage{graphicx}
\usepackage{txfonts}
\newcommand{\etal}{et\,al.\ }
\newcommand{\logg}{\mbox{$\log g$}}
\newcommand{\Teff}{\mbox{$T_\mathrm{eff}$}}

\newcommand{\rxj}{\object{RX\,J2117.1$+$3412}}

\newcommand{\keins}{\object{K1$-$16}}

\begin{document}
   \title
   {Fluorine in extremely hot post-AGB stars: evidence for nucleosynthesis\thanks
{Based on observations made with the NASA-CNES-CSA Far Ultraviolet 
Spectroscopic Explorer. FUSE is operated for NASA by the Johns Hopkins
University under NASA contract NAS5-32985.}
   }
 
   \author{K. Werner$^1$, T. Rauch$^{1,2}$ \and J.W. Kruk$^3$}
   \offprints{K. Werner}
   \mail{werner@astro.uni-tuebingen.de}
 
   \institute
    {
     Institut f\"ur Astronomie und Astrophysik, Universit\"at T\"ubingen, Sand 1, 72076 T\"ubingen, Germany
\and
Dr.-Remeis-Sternwarte, Universit\"at Erlangen-N\"urnberg, Sternwartstra\ss e 7, 96049 Bamberg, Germany
\and
Department of Physics and Astronomy, Johns Hopkins University,
Baltimore, MD 21218, U.S.A.
}
    \date{Received xxx / Accepted xxx}
   \authorrunning{K. Werner et al.}
   \titlerunning{Fluorine in extremely hot post-AGB stars}
   \abstract{We have discovered lines of highly ionized fluorine (\ion{F}{v} and
\ion{F}{vi}) in the far-UV spectra of extremely hot
(\Teff=85\,000--150\,000\,K) post-AGB stars. Our sample comprises H-rich
central stars of planetary nebulae as well as H-deficient PG1159 stars. We
performed non-LTE calculations and find strong F overabundances (up to 10$^{-4}$
by mass, i.e., 250 times solar) in a number of PG1159 stars, while F is
essentially solar in the H-rich stars. Since PG1159 stars are believed to
exhibit intershell matter of the preceding AGB phase on their surface, their
chemical analyses allow for a direct insight into nucleosynthesis processes
during the AGB phase. The high F abundances in PG1159 stars confirm the
conclusion from abundance determinations in giants, that F is synthesized in AGB
stars and that the F enrichment in the intershell must be very high.
        }
   \maketitle

\begin{figure}[tbp]
\includegraphics[width=\columnwidth]{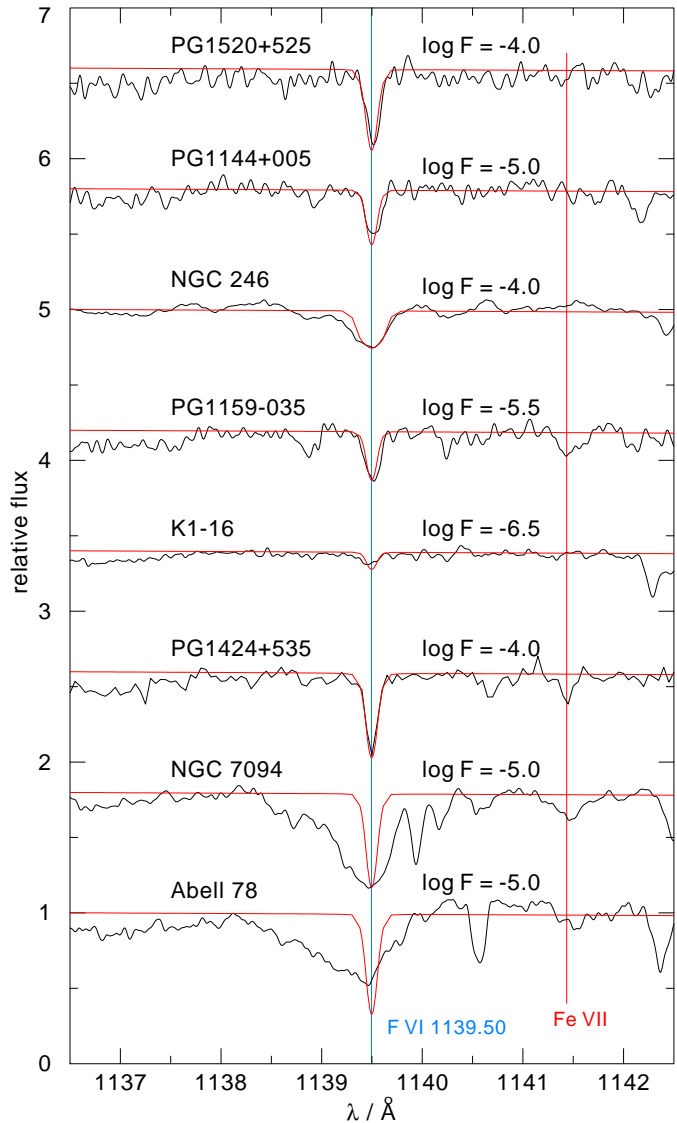}
  \caption[]{
Identification of the \ion{F}{vi}~1139.5\AA\ line in PG1159 stars. The F
abundances of the overplotted models are given (the solar abundance is
log\,F=$-6.4$). The observed line profiles of NGC\,7094 and Abell~78 are
significantly broader than the computed ones due to wind effects, which are
neglected in our models. Rotational broadening (45\,km/s) is applied to the
NGC\,246 profile. The position of a \ion{Fe}{vii} line is also marked. It is
very weak or not detectable in the PG1159 stars because of their iron-deficiency
(Miksa \etal 2002). Other lines are of interstellar origin.}
  \label{fig_h-def}
\end{figure}

\begin{figure}[tbp]
\includegraphics[width=\columnwidth]{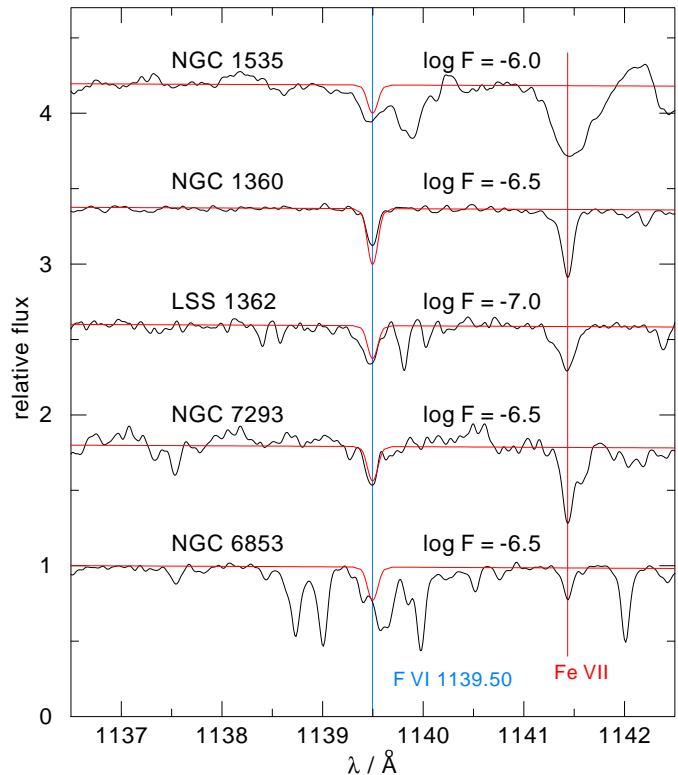}
  \caption[]{
Like Fig.\,\ref{fig_h-def}, but here for the hydrogen-rich central stars. The
finally adopted F abundances are listed in Tab.\,\ref{H-rich_objects_tab}. The
\ion{Fe}{vii} line is strongly detected and probably consistent with solar Fe
abundance (Hoffmann \etal 2005). It is not included in the models being plotted.}
  \label{fig_h-rich}
\end{figure}

\begin{figure}[tbp]
\includegraphics[width=\columnwidth]{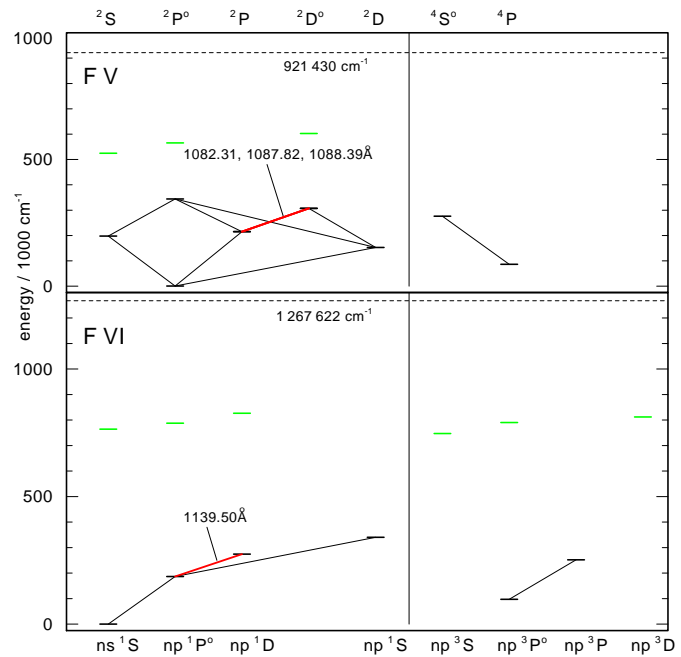}
  \caption[]{
Grotrian diagrams of our \ion{F}{v} and \ion{F}{vi} model ions. All levels
linked by line transitions are computed in NLTE, others are LTE
levels. Transitions observed in the FUSE spectra are marked. All other lines are
located in the EUV, inaccessible with FUSE.  }
  \label{fig_grotrian}
\end{figure}

\section{Introduction}

For a long time, the solar system was the only location of the Galaxy with known
fluorine ($^{19}$F) abundance (see e.g.\ Meynet \& Arnould 2000). The production
site of F has always been a puzzle,  although its abundance is very small
(log\,F\,=$-6.4$ by mass; Grevesse \& Sauval 1998).  This situation changed
since the detection of F overabundances (up to 30 times solar) in red giants
(Jorissen \etal 1992) with the discovery of an infrared HF molecule line. This
proves that AGB stars are in fact F producers, but the question is still open as
to what extent AGB stars contribute to the solar-system and Galactic F
content. A recent analysis of HF lines in giants in $\omega$\,Cen, though,
suggests that AGB stars are not the main F contributor (Cunha \etal 2003).

Other potential F producers are Wolf-Rayet stars which increase the Galactic F
content by wind mass-loss (Meynet \& Arnould 2000), and in Type~II supernovae
$^{19}$F is created by spallation of $^{20}$Ne by neutrinos (Woosley \& Weaver
1995). Renda \etal (2004) address the relative importance of these three
primary astrophysical factories of fluorine production in the Galaxy.

The general problem of fluorine production is that $^{19}$F, the only stable F
isotope, is rather fragile and readily destroyed in hot stellar interiors by
hydrogen via $^{19}$F(p,$\alpha$)$^{16}$O and helium via
$^{19}$F($\alpha$,p)$^{22}$Ne. The nucleosynthesis path for F production in
He-burning environments of AGB and Wolf-Rayet stars is:

$^{14}$N($\alpha$,$\gamma$)$^{18}$F($\beta^+$)$^{18}$O(p,$\alpha$)$^{15}$N($\alpha$,$\gamma$)$^{19}$F.

\noindent
Protons are provided by the $^{14}$N(n,p)$^{14}$C reaction with neutrons
liberated from $^{13}$C($\alpha$,n)$^{16}$O. The $^{14}$N and $^{13}$C
nuclei can result from ashes of hydrogen burning  by CNO cycling. By this,
$^{19}$F is enriched in the convective He-rich intershell of AGB stars and then
dredged-up to the surface (3rd dredge up). However, if there is no other
source of $^{14}$N and $^{13}$C nuclei than the ashes of hydrogen burning, then
$^{19}$F will be produced in a negligible amount only. To yield the observed
overabundances, primary $^{13}$C and $^{14}$N nuclei are an absolute
necessity. They probably result from proton mixing (``partial mixing'') into the
He-rich intershell, the same mixing necessary to activate the
s-process in those stars (Mowlavi \etal 1998).

Current stellar models cannot match the highest F abundances observed in AGB
stars (Lugaro \etal 2004). The problems in modeling arise from the mixing
and burning processes during thermal pulses, and still uncertain nuclear
reaction rates, particularly for a competing process that prevents the F
synthesis: the $^{15}$N(p,$\alpha$)$^{12}$C reaction removes $^{15}$N and
protons from the F production chain. In order to explain the highest observed F
enrichments, one would require very large F overabundances in the
intershell($10^{-3}$--$10^{-4}$, i.e.\ 250--2500 times solar). The models do
produce a fluorine overabundance, but not high enough.

In this paper we announce the detection of F in extremely hot post-AGB stars,
which gives new insight into the problem and strongly confirms the role of AGB
stars as Galactic F producers. We derive very high F overabundances in the
hydrogen-deficient stars of our sample, the so-called PG1159 stars. This is of
particular interest, because these objects are thought to exhibit intershell
abundances on their surface, probably as a consequence of a late He-shell flash
(Werner 2001). An analysis of a representative subset of our full sample, for
which we have spectra with particularly good S/N, will be presented  here.

We are currently analysing a large sample of very hot post-AGB stars in order to
derive atmospheric parameters and to gain insight into the evolutionary history
of these stars. In particular we are interested in the PG1159 stars because many
aspects of their evolution are still unclear. UV spectroscopy with the Far
Ultraviolet Spectroscopic Explorer (FUSE) is most suitable for this work because
all metals in the hot photospheres are highly ionised. During the course of our
analyses we realized that, although the spectra are poor in the number of line
features (from highly ionized species, essentially He, C, O, Ne), a number of
lines still cannot be identified.  One of these features, located at 1139.5\AA,
is detectable in most PG1159 stars, over a quite large temperature range
(\Teff=110\,000\,K--150\,000\,K). It is sometimes very prominent with central
depressions of 50\%, and it is also seen in many H-rich central stars with
temperatures ranging between 70\,000--130\,000\,K. We identify this line as due
to \ion{F}{vi}, which is confirmed by the additional identification of a
\ion{F}{v} multiplet in the  ``coolest'' PG1159 stars and H-rich central stars
of our sample.

\begin{table*}
\caption{Parameters of the H-deficient program stars. 
Abundances are given in mass fractions. The solar F abundance is log\,F=$-6.4$. We
note that in some objects trace nitrogen (abundance of order 0.01) was found
(see last column). Ne abundances from Werner \etal
(2004a).\label{H-poor_objects_tab}}
\begin{tabular}{l r r r r r r r r r l l}
      \hline
      \hline
      \noalign{\smallskip}
Object  & \Teff & \logg & H & He  & C   & O   & Ne  &log F  & mass       & ref.&remark\\
        & [kK]  & (cgs) &   &     &     &     &     &       &[M$_\odot$] &     &\\
      \noalign{\smallskip}
      \hline
      \noalign{\smallskip}
H1504+65    & 200 & 8.0 &   &$<$.01&.48 & .48 & .02 &$<-4.5$& 0.84 & A& Mg=0.02\\
\rxj        & 170 & 6.0 &   & .38 & .54 & .06 & .02 &$<-5.0$  & 0.70 & B&\\
PG1520+525  & 150 & 7.5 &   & .43 & .38 & .17 & .02 & $-4.0$  & 0.65 & J&\\
PG1144+005  & 150 & 6.5 &   & .38 & .58 & .02 & .02 & $-5.0$  & 0.57 & C& trace nitrogen\\
NGC\,246    & 150 & 5.7 &   & .62 & .30 & .06 & .02 & $-4.0$  & 0.76 & H&\\
PG1159-035  & 140 & 7.0 &   & .33 & .48 & .17 & .02 &$-5.5$ & 0.54 & I,J&\\
\keins      & 140 & 6.4 &   & .33 & .48 & .17 & .02 &$-6.5$ & 0.59 & G&\\   
HS2324+397  & 130 & 6.2 &.21& .41 & .37 & .01 &     &$<-6.0$  & 0.59 & E& poor S/N, strong i.s. H$_2$ contamination\\
Longmore 4  & 120 & 5.5 &   & .45 & .42 & .11 & .02 &$<-6.5$& 0.65 & D&\\
PG1424+535  & 110 & 7.0 &   & .49 & .43 & .06 & .02 & $-4.0$  & 0.50 & J&\\
NGC\,7094   & 110 & 5.7 &.35& .41 & .21 & .01 & .02 &$-5.0$   & 0.59 & E& 1139.5\AA\ wind profile\\      
Abell 78    & 110 & 5.5 &   & .35 & .50 & .15 &     &$-5.0$   & 0.77 & F& 1139.5\AA\ wind profile; trace nitrogen\\
PG1707+427  &  85 & 7.5 &   & .43 & .38 & .17 & .02 &$-4.0$   & 0.54 & J& poor S/N; trace nitrogen\\
 \noalign{\smallskip}
      \hline
     \end{tabular}
\\References: 
A: Werner \etal 2004b,
B: Werner \etal 1996,
C: Werner \& Heber 1991,
D: Werner \etal 1992,
E: Dreizler \etal 1997,
F: Koesterke \& Werner 1998,
G: Kruk \& Werner 1998,
H: Rauch \& Werner 1997,
I: Werner \etal 1991,
J: Dreizler \& Heber 1998
\end{table*}

\section{Fluorine abundance analysis}\label{analy}

FUSE observations and data reduction for most of our program stars were
described in a previous work (Werner \etal 2004a). Further observations utilized
here will be described in the comprehensive work mentioned above.

We have designed a fluorine model atom for NLTE line formation
calculations. These are performed using and keeping fixed the physical structure
(temperature, densities) of line blanketed NLTE model atmospheres which are
described in detail in Werner \etal (2004a). In short, they are plane-parallel
and in hydrostatic and radiative equilibrium. For the PG1159 stars, the model
parameters and references for the previous analyses are given in
Table~\ref{H-poor_objects_tab}. The models are composed of H, He, C, O, and
Ne. For the H-rich central stars we calculated models with a solar composition,
which is close but not identical to the literature values given in
Table~\ref{H-rich_objects_tab} (the model for NGC\,1360 accounts for the
oversolar He abundance). Due to convergence problems, the models for NGC\,1360
and NGC\,1535 do not include O and Ne. Both restrictions are not expected to
have a strong influence on the calculated fluorine line profiles.

For each star we performed F calculations with different abundances in order to
estimate the observed abundance and the error of the analysis. For two objects
our static models are not adequate. NGC\,7094 and Abell~78 show distinct
extended blue wings in the \ion{F}{vi} profile (Fig.\,\ref{fig_h-def}), which
clearly are the signature of mass-loss. Both stars also show strong P~Cyg
profiles in the \ion{C}{iii} and \ion{O}{vi} resonance lines. Modeling of
expanding atmospheres is beyond the scope of this paper and we assign a larger
error to the F abundance determination in these two cases.

The model atom considers ionization stages \ion{F}{iv}--\ion{F}{viii},
represented by 2, 8, 6, 2, 1 NLTE levels, respectively, plus a number of LTE
levels (Fig.\,\ref{fig_grotrian}). In the ions \ion{F}{v}--\ion{F}{vii} we
include 9, 4, 1 line transitions. Atomic data were taken from the NIST and
Opacity Project (OP, Seaton \etal 1994) databases. Special attention was given
to the lines which we compare to the observation. In order to compute the three
profiles of the \ion{F}{v} multiplet ($\lambda\lambda$ 1082.31\AA, 1087.82\AA,
1088.39\AA), we performed fine structure splitting of energy levels and
distributed level population densities according to statistical weights. For all
lines we assumed quadratic Stark broadening for the profile
calculation. Oscillator strengths are taken from OP.

It is remarkable that these \ion{F}{v} and \ion{F}{vi} lines are detectable in
the observations, even at very low F abundances. This is because they are the
only UV lines that arise from relatively low excited levels (excitation energy
about 200\,000\,cm$^{-1}$). All other strong lines from these ions, arising from
such low lying levels, are located in the inaccessible EUV spectral region.

In seven out of twelve PG1159 stars we identified the \ion{F}{vi}~1139.5\AA\
line. They are plotted in Fig.\,\ref{fig_h-def} together with synthetic profiles
from those grid models with F abundances that most closely match the
observations. In the two hottest PG1159 stars (H1504+65 and \rxj)
\ion{F}{vi}~1139.5\AA\ is not detected because the ionization shifts away from
\ion{F}{vi} at these high temperatures, hence, only large upper limits for the
abundance can be given. The non-detection of this line in Longmore~4 and its
weakness in \keins\ suggest an F abundance not higher than solar. The result for
HS2324+397 is not very strict because of poor S/N of the spectra and a strong
contamination of the \ion{F}{v} multiplet by interstellar H$_2$
lines. Fig.\,\ref{fig_h-rich} displays the spectra of all five H-rich central
stars from our sample. The \ion{F}{vi}~1139.5\AA\  line profile for NGC\,6853 is
unique in being contaminated by warm H$_2$ along the line of sight.  The
\ion{F}{vi} line is bracketed by weak H$_2$  features that account for at most
half of the observed absorption on either side of the \ion{F}{vi} line.

\begin{figure}[tbp]
\includegraphics[width=\columnwidth]{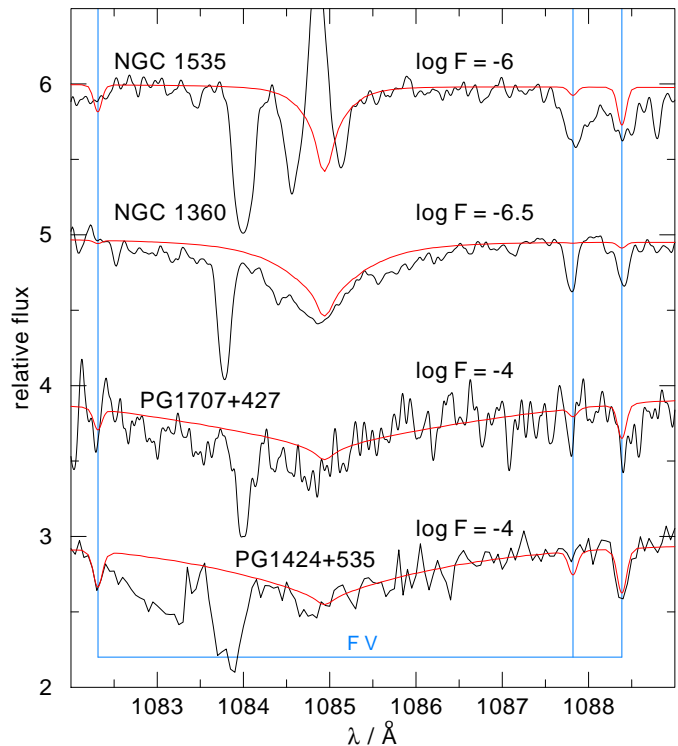}
  \caption[]{
Identification of a \ion{F}{v} multiplet in two H-rich central stars (upper two
spectra) and two PG1159 stars. The broad absorption line
centered around 1085\AA\ is from \ion{He}{ii}. The deep asymmetric depression
over 1082--1083.5\AA\ in PG\,1424+535 is spurious.
}
  \label{fig_fv}
\end{figure}

Fig.\,\ref{fig_fv} shows those four spectra in which we could identify the
\ion{F}{v} multiplet. They are from two H-rich central stars and from two PG1159
stars, all of which are relatively ``cool'' objects, for which the ionization
balance shifts from \ion{F}{vi} towards \ion{F}{v} when compared to the hotter
objects. All three lines of this multiplet are well matched by the model for
PG1424+535 with the F abundance kept fixed from the fit to the \ion{F}{vi} line,
whereas the spectrum for PG1707+427 appears too noisy for a quantitative
comparison. In contrast, the two computed \ion{F}{v} lines at 1087.82\AA\ and
1088.39\AA\ for the two H-rich stars (NGC\,1535 and NGC\,1360) appear much too
weak compared to the observation. But the third line component, at 1082.31\AA,
is not detectable in these stars, which is in agreement with the computed weak
profiles. According to the models, the components at 1082.31\AA\ and 1088.39\AA\
should be similar in strength as is the case with PG1424+535. The reason for
the unexpectedly strong components at 1087.82\AA\, 1088.39\AA\ in NGC\,1535 and
NGC\,1360 remains unexplained.

The F abundance in the models was varied in steps of 0.5\,dex.  The derived F
abundances are given in Tables~\ref{H-poor_objects_tab} and
\ref{H-rich_objects_tab}.  We estimate the error of our abundance analysis to
0.5\,dex, taking into account possible errors in the values for \Teff\ and
\logg, which are of the order 10\% and 0.3\,dex, respectively. For NGC\,7094 and
Abell~78 we must accept larger errors, maybe 1~dex, for the reasons mentioned
above.

\section{Results and discussion}

\begin{figure}[tbp]
\includegraphics[width=\columnwidth]{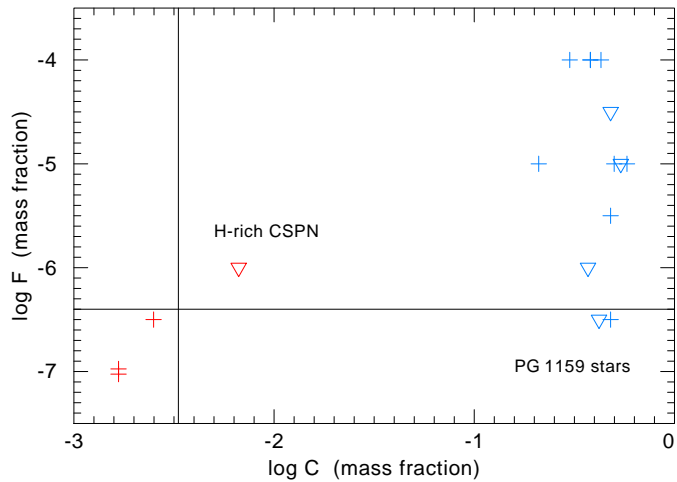}
  \caption[]{
The fluorine abundance as a function of the carbon abundance in our program
stars (except for NGC\,1535, for which no C abundance is available). 
The C-rich PG1159 stars are in general fluorine enriched in contrast to
the (essentially) C-normal central stars. The vertical and horizontal lines
denote the solar C and F abundance values, respectively. Typical errors for
abundances are 0.3--0.5 dex. Triangles denote stars for which only upper limits
for F could be found.
}
  \label{fig_f_vs_c_abundance}
\end{figure}

Our fluorine abundance analysis can be summarized as follows
(Fig.\,\ref{fig_f_vs_c_abundance}).  We have identified fluorine lines in five
H-rich central stars and in eight H-deficient PG1159 stars. Within error limits,
the hydrogen-rich central stars have solar F abundances, i.e.\
log\,F\,=\,$-6.4$. Most of the PG1159 stars show extreme F overabundances,
ranging from 10--250 times solar (log\,F\,=\,$-5.4, ... ,-4$). There is one
PG1159 star with a solar F abundance (\keins) and another one with an upper
limit of solar abundance (Longmore~4). By comparing the CSPN (which have
essentially solar C abundances) and the C-rich PG1159 stars we find that the
general trend of increasing F abundance with increasing C abundance is in
accordance with nucleosynthesis modeling (Goriely \& Mowlavi 2000). But it must
be stated that appropriate modeling for PG1159 stars, i.e.\ taking into account
a late He-shell flash, is still lacking.

Concerning the H-rich stars we can conclude that the original F content in
the stellar envelopes remains essentially unchanged. No enrichment due to
dredge-up of F, which might have been synthesized in the AGB star, has
occurred. Jorissen \etal (1992) found in their AGB star sample that the F
enrichment is correlated with the enrichment of C. Matching to this, our five
central stars are not significantly C-rich, see Table~\ref{H-rich_objects_tab}
where we list the C/O ratio determined in a recent analysis of FUSE and HST
spectra (Traulsen \etal 2004). The mass range 0.56--0.65\,M\,$_\odot$,
covered by our CSPN, corresponds to an initial-mass range of
1.2--2.6\,M\,$_\odot$ (Weidemann 2000). According to the model calculations of
Lugaro \etal (2004), a significant F enrichment on the AGB star surface can only
be expected for stars with an inital mass larger than about
2.5\,M\,$_\odot$. This is in accordance with our abundance determinations. The
iron abundance in the H-rich stars appears to be solar (Hoffmann \etal 2005),
which is consistent with these findings.

A truly remarkable result is the high F overabundance found in the PG1159
stars. This strongly corroborates the idea, that these stars show intershell
matter of the precursor AGB star on their surface. Herwig \etal (1999) were able
to quantitatively explain the surface abundances of He, C, O, and Ne as a
consequence of a late He-shell flash. According to AGB stellar models, F is also
strongly enriched in the intershell, so that the detected overabundances in the
PG1159 stellar spectra fit qualitatively well into the overall picture.
The detected deficiency of iron in PG1159 stars (Miksa \etal 2002) can also be
understood. Neutron captures in the s-processing environments transform Fe into
heavier elements and deplete the Fe content in the He-rich intershell. Hence,
dredge-up of that intershell matter resulting from the late He-shell flash
explains both the observed iron deficiency and the fluorine enrichment in the
atmospheres of PG1159 stars.

What are the implications from the PG1159 star analyses for the modeling of
nucleosynthesis in thermally pulsing AGB stars? The models of Lugaro \etal
(2004) predict that the F enrichment in the intershell is a strong function of
the initial stellar mass. For solar metallicity models, it is maximal at
$M$=3\,M$_\odot$, with log\,F\,=\,$-4.2$. It drops to log\,F\,=\,$-5$ for
$M$=5\,M$_\odot$, and to log\,F\,=\,$-6.2$ for a $M$=6.5\,M$_\odot$ star. This
variety of possible F intershell abundances is in agreement with our results for
the PG1159 stars. Even \keins\ with a solar F abundance could be explained as
the descendant of a particularly massive star, say 7-8\,M$_\odot$, however, the
spectroscopically determined mass ($M$=0.59\,M$_\odot$) points to a less massive
progenitor ($M$=2\,M$_\odot$, using the initial-final-mass relation of Weidemann
2000). Generally, we do not see a trend of the F abundance in PG1159 stars
with the initial stellar mass (our sample covers the range
$M_i$=0.8--4.5\,M$_\odot$).

Hence, aside from this particular problem, the F intershell abundances in the
Lugaro \etal (2004) models are large enough to explain even the highest
overabundances observed in PG1159 stars. In contrast, these authors conclude
that their intershell abundances are not high enough to explain the F
overabundances found in AGB stars. When mixed from the intershell into the
H-rich convective surface layer, F gets dredged up to the surface, but is too
diluted to match the large overabundances reported by Jorissen \etal (1992).

Although our results seem to confirm that the F enrichment in the intershell is
described correctly by stellar models, this may be a chance coincidence. Without
proper modeling including a detailed nuclear network it is not clear how the F
abundance is changed during the event of a late helium-shell flash. Further work
in this direction is urgently needed.

\begin{table}
\caption{Parameters of the H-rich stars. 
Number abundance ratios of H, He, C, and O are given. The F abundance is in mass
fraction; the solar abundance is log\,F=$-6.4$. Photospheric and stellar
parameters are from Traulsen \etal (2005), except for NGC\,1535 (M\'endez \etal
1992).\label{H-rich_objects_tab}}
\begin{tabular}{l r r r r r r r r r l l}
      \hline
      \hline
      \noalign{\smallskip}
Object  & \Teff & \logg & He/H & C/O & log F  & mass       \\
        & [kK]  & (cgs) &      &     &        &[M$_\odot$] \\
      \noalign{\smallskip}
      \hline
      \noalign{\smallskip}
NGC\,1535 &  70 & 4.6   & 0.12 &     & $-6.5$ & 0.65 \\
NGC\,1360 &  97 & 5.3   & 0.25 & 1.0 &  $-7.0$ & 0.65 \\   
LSS\,1362 & 114 & 5.7   & 0.10 & 1.0 &  $-7.0$ & 0.60 \\
NGC\,7293 & 120 & 6.3   & 0.03 & .86 & $-6.5$ & 0.56 \\
NGC\,6853 & 126 & 6.5   & 0.10 & 2.0 &$<-6.0$ & 0.58 \\
 \noalign{\smallskip}
      \hline
     \end{tabular}
\end{table}

\begin{acknowledgements}
UV data analysis in T\"ubingen is supported by the DLR under grant 50\,OR\,0201.
JWK is supported by the FUSE project, funded by NASA contract NAS5-32985. We
thank Falk Herwig (LANL) for helpful discussions on evolutionary aspects, and
Alexander Kramida (NIST) for his advice on atomic data. We thank the referee,
A.~Jorissen, for constructive remarks.
\end{acknowledgements}


\begin{thebibliography}{}

\bibitem{Cun03} Cunha, K., Smith, V.~V., Lambert, D.~L., \& Hinkle, K.~H. 2003,
  AJ, 126, 1305

\bibitem{DH98} Dreizler, S., \& Heber, U. 1998, A\&A, 334, 618

\bibitem{Dr97} Dreizler, S., Werner, K., \& Heber, U. 1997, in Planetary Nebulae,
                ed.\ H.J.\,Habing, H.J.G.L.M.\,Lamers, IAU Symp. 180, Kluwer,
                p.\,103

\bibitem{Gor00} Goriely, S., \& Mowlavi, N. 2000, A\&A, 362, 599

\bibitem{Gre98} Grevesse, N., \& Sauval, A.~J. 1998, Space Sci.\ Rev.\ 85, 161

\bibitem{Her99} Herwig, F., Bl\"ocker, T., Langer, N., \& Driebe, T. 1999, A\&A,
  349, L5

\bibitem{Hof05} Hoffmann, A.~I.~D., Traulsen, I., Dreizler, S., Rauch, T., Werner,
  K., \& Kruk, J.~W. 2005, in White Dwarfs, eds.\
  D.\,Koester, S.\,Moehler, ASP Conf.\ Series, in press 

\bibitem{Jor92} Jorissen, A., Smith, V.~V., \& Lambert, D.~L. 1992, A\&A, 261, 164

\bibitem{Koe98} Koesterke, L., \& Werner, K. 1998, ApJL, 500, 55

\bibitem{KW98} Kruk, J.~W., \& Werner, K. 1998, ApJ, 502, 858

\bibitem{Lug04} Lugaro, M., Ugalde, C., \& Karakas, A.~I., \etal 2004, ApJ, 615, L934

\bibitem{Men92} M\'endez, R.~H., Kudritzki, R.~P., \& Herrero, A. 1992, A\&A, 260, 329

\bibitem{Mey00} Meynet, G., \& Arnould, M. 2000, A\&A, 355, 176

\bibitem{Mi02} Miksa, S., Deetjen, J.~L., Dreizler, S., Kruk, J., Rauch, T., \&
	Werner, K. 2002, A\&A, 389, 953

\bibitem{Mow98} Mowlavi, N., Jorissen, A., \& Arnould, M. 1998, A\&A, 334, 153

\bibitem{RW97} Rauch, T., \& Werner, K. 1997, in The Third Conference on Faint
        Blue Stars, ed.\ A.G.D.\,Philip, J.\,Liebert, R.A.\,Saffer, (L.\,Davis
        Press, Schenectady, NY), p.\,217

\bibitem{Ren04} Renda, A., Fenner, Y., Gibson, B.~K., \etal 2004, MNRAS, 354, 575

\bibitem{Sea94} Seaton, M.~J., Yan, Y., Mihalas, D., \& Pradhan, A.~K. 1994, MNRAS, 266, 805

\bibitem{Tra04} Traulsen, I., Hoffmann, A.~I.~D., Dreizler, S., Rauch, T., Werner,
  K., \& Kruk, J.~W. 2005, in White Dwarfs, eds.\
  D.\,Koester, S.\,Moehler, ASP Conf.\ Series, in press

\bibitem{Wei00} Weidemann, V. 2000, A\&A 363, 647

\bibitem{KW01} Werner, K. 2001, in Low Mass Wolf-Rayet Stars:
	Origin and Evolution, eds. T.\,Bl\"ocker, L.B.F.M.\,Waters,
	A.A.\,Zijlstra, Ap\&SS, 275, 27

\bibitem{WH91} Werner, K., \& Heber, U. 1991, A\&A, 247, 476

\bibitem{Wetal91} Werner, K., Heber, U., \& Hunger, K. 1991, A\&A, 244, 437

\bibitem{Wetal92} Werner, K., Hamann, W.-R., Heber, U., Napiwotzki, R., Rauch,
        T., \& Wessolowski, U. 1992, A\&A, 259, L69

\bibitem{W96} Werner, K., Dreizler, S., Heber, U., \etal 1996, A\&A,
  307, 860

\bibitem{W04a} Werner, K., Rauch, T., Reiff, E., Kruk, J.~W., \& Napiwotzki,
  R. 2004a, A\&A, 427, 685

\bibitem{W04b} Werner, K., Rauch, T., Barstow, M.~A., \& Kruk, J.~W. 2004b, A\&A, 421, 1169

\bibitem{Woo95} Woosley, S.~E., \& Weaver, T.~A. 1995, ApJS, 101, 181

\end{thebibliography}
\end{document}